\documentclass[10pt,twocolumn,amsmath,amssymb,aps,floatfix,preprintnumbers,nofootinbib,nobibnotes,bibnotes]{revtex4-2}
\usepackage{graphicx}
\usepackage{dcolumn}
\usepackage{bm}
\usepackage{hyperref}
\usepackage[mathlines]{lineno}
\usepackage{epsfig}  
\usepackage{color}
\usepackage{slashed}
\usepackage{float}
\usepackage[table]{xcolor}
\usepackage{amsmath}
\usepackage{soul}

\large

\begin{document}

\title{Genuine lepton-flavor-universality-violating observables in the $\tau-\mu$ sector of $B \to (K,\,K^*) \ell \ell $ decays}

\author{Ashutosh Kumar Alok}
\email{akalok@iitj.ac.in}
\affiliation{Indian Institute of Technology Jodhpur, Jodhpur 342037, India}

\author{Neetu Raj Singh Chundawat}
\email{chundawat.1@iitj.ac.in}
\affiliation{Indian Institute of Technology Jodhpur, Jodhpur 342037, India}

\author{Jitendra Kumar}
\email{jkumar@iitj.ac.in}
\affiliation{Indian Institute of Technology Jodhpur, Jodhpur 342037, India}

\author{Arindam Mandal}
\email{mandal.3@iitj.ac.in}
\affiliation{Indian Institute of Technology Jodhpur, Jodhpur 342037, India}

\author{Umberto Tamponi}
\email{tamponi@to.infn.it}
\affiliation{INFN Sezione di Torino, Via Pietro Giuria 1, I-10125 Torino, Italy}

\begin{abstract}
 It was previously shown that unlike the ratios $R_K^{\mu e} \equiv R_K \equiv \Gamma(B \to K \mu^+ \mu^-)/\Gamma(B \to K e^+ e^-)$ and $R_{K^*}^{\mu e} \equiv R_{K^*} \equiv \Gamma(B \to K^* \mu^+ \mu^-)/\Gamma(B \to K^* e^+ e^-)$, the ratios $R_K^{\tau \mu}$ and $R_{K^*}^{\tau \mu}$ can deviate from their Standard Model (SM) predictions even with universal new physics couplings. This observation highlights the critical need to identify and establish genuine lepton flavor universality violating (LFUV) observables in the $\tau-\mu$ sector. This work embarks on establishing genuine LFUV ratio observables in \(B \to K \ell \ell\) and \(B \to K^* \ell \ell\) decays through comprehensive analysis of their angular distributions. We find that like $R_{K^*}^{\tau \mu}$, the ratios $R_{A_{FB}}^{\tau \mu}$ and $R_{f_L}^{\tau \mu}$ do not qualify as genuine LFUV observables, whereas the ratios of all optimized observables in $B \to K^* \ell \ell$ decays within the $\tau-\mu$ sector definitively do. In the case of $B \to K \ell \ell$ decays, similar to $R_K^{\tau \mu}$, the ratio $R_{F_H}$ is influenced by mass effects and therefore cannot be considered a genuine LFUV observable in the $\tau-\mu$ sector. However, the ratio $\Gamma_\tau(1-F_{H}^{\tau})/\Gamma_\mu(1-F_{H}^{\mu})$ stands as the sole genuine LFUV observable in  $B \to K \ell \ell$  decays. Furthermore, by making use of new physics Lorentz structures which provide a better fit to the current $b \to s \ell \ell$ data as compared to the SM, we demonstrate how the non-genuine LFUV ratios $R_{A_{FB}}^{\tau \mu}$ and $R_{f_L}^{\tau \mu}$ can be employed to distinguish between framework with solely universal lepton couplings and those with both universal and non-universal couplings.
\end{abstract}

\maketitle

\section{Introduction} 

 The Large Hadron Collider (LHC) has yet to unveil any particles beyond those predicted by the Standard Model (SM) of electroweak interactions. This absence may be attributed to the possibility that these new particles are massive enough to not yield sufficient production rates at the current LHC energies. In light of this, the significance of ongoing experiments like LHCb and Belle II has notably escalated, given their capacity to explore new physics at scales much higher than those attainable by direct search experiments such as ATLAS and CMS, thanks to their ability to make precise measurements of the properties of rare $B$ meson decays. 

The decay induced by the quark level transition $b \to s \ell \ell$ has immense potential to probe new physics. This decay mode is highly suppressed in SM and originates several different final states such as $B \to X_s \ell \ell$,  $B \to (K,\,K^*) \ell \ell$, $B_s \to \phi \ell \ell$ and $B_s \to \ell \ell$
and hence providing a number of avenues to hunt for physics beyond SM. Moreover, the SM $CP$ violating effects in this sector are too small to be detected at the current collider facilities. Hence any measurement of such effects will provide an unambiguous signature of new physics\cite{Fleischer:2022klb,Fleischer:2023zeo,Fleischer:2024fkm,SinghChundawat:2022zdf,Gangal:2022ole,Das:2022xjg,Geng:2022pld}. 

There are already a plethora of measurements of several observables sensitive to new physics in a number of decays involving the quark level transitions $b \to s e^+ e^-$ and $b \to s \mu^+ \mu^-$. Apart from the measurements of the branching ratios of  $B \to X_s \ell \ell $,  $B \to (K,\,K^*) \ell \ell$ $(\ell=e,\,\mu)$, $B_s \to \phi \mu^+ \mu^-$ and $B_s \to \mu^+ \mu^-$ decays,  a number of angular observables in $B \to K^* e^+ e^-$,  $B \to K^* \mu^+ \mu^-$ and $B_s \to \phi \mu^+ \mu^-$ decays have also been measured in several $q^2$ bins.  A few of these measurements do not agree with the predictions of the SM. The most striking discrepancy lies in the measurement of the branching ratio of $B_s \to \phi \mu^+ \mu^-$ decay. Specifically, the branching ratio of this decay mode in the (1-6) $q^2$ bin disagrees with the SM prediction at the level of 3.5$\sigma$ \cite{bsphilhc3}.

In addition to the observables mentioned above, which are either related to electron or muon channels, there are observables that are related to both channels. These observables have the potential to test the violation of Lepton Flavor Universality (LFU), which is a fundamental aspect deeply ingrained in the symmetry structure of the SM. The ratios $R_K^{\mu e}\equiv R_K\equiv\Gamma(B \to K \mu^+  \mu^-)/\Gamma(B \to K e^+ e^-)$ and $R_{K^*}^{\mu e}\equiv R_{K^*}\equiv\Gamma(B \to K^* \mu^+  \mu^-)/\Gamma(B \to K^* e^+ e^-)$ serve as pivotal observables for probing LFU violation (LFUV) within the $\mu-e$ sector in $B \to K \ell \ell$ and $B \to K^* \ell \ell$ decays, respectively. This is due to their unique capability to depart from SM predictions solely in the presence of non-universal new physics, where the couplings differ between the $\mu-e$ sector in the $b \to s \ell \ell$ decay process. Consequently, any significant deviation from their SM expectations would not only corroborate the existence of new physics but also confirm its non-universal nature. 
It should be noted that the current measurements of $R_K^{\mu e}$ and $R_{K^*}^{\mu e}$  are consistent with their SM predictions \cite{LHCb:2022qnv,LHCb:2022zom,Bordone:2016gaq,Hiller:2003js,Isidori:2020acz,Isidori:2022bzw,Nabeebaccus:2022pje}.

The ratios $R_K^{\tau \mu}$ and $R_{K^*}^{\tau \mu}$ were anticipated to serve a similar role in the $\tau-\mu$ sector, suggesting that any observed deviation from SM predictions in these observables would indicate LFUV type of new physics in the $\tau-\mu$ sector. Even though the current measurements of 
$R_{K^{(*)}}^{\mu e}$ are consistent with their SM predictions, the global analysis of $b \to s \ell \ell$ ($\ell=e,\,\mu$ data (including $R_{K^{(*)}}^{\mu e}$ measurements) does not rule out the possibility of having a moderate values of LFUV components of new physics couplings in a number of new physics scenarios \cite{SinghChundawat:2022ldm,Alguero:2023jeh}. Thus, the existing data does not eliminate the possibility of LFU violation in the $\tau-\mu$ sector \cite{SinghChundawat:2022ldm}.

However, contrary to expectations, it was demonstrated in \cite{Alok:2023yzg} that these ratios can deviate from their SM predictions even when the new physics couplings are universal. This deviation was linked to mass-related effects associated with the involvement of $\tau$ and $\mu$ leptons.

In this study, we undertake the task of identifying and constructing genuine LFUV  ratio observables within the $\tau-\mu$ sector in $B \to K \ell \ell$ and $B \to K^* \ell \ell$ decays free from mass-related effects, through an analysis of the full angular distribution of these decays. We also illustrate how non-genuine LFUV observables can be utilized to distinguish between universal and non-universal types of new physics frameworks.  For this, we consider new physics scenarios which provide a better fit to the current $b \to s \ell \ell$ data as compared to the SM.   

The plan of the work is as follows. In Sec. \ref{for-tmu}, we discuss the formalism, which includes the $b \to s \ell \ell$ effective Hamiltonian along with constraints on the new physics Wilson Coefficients (WCs). In Sec. \ref{obs-bks}, we construct genuine LFUV observables in $\tau-\mu$ sector in $B \to K^*  \ell \ell$ decays. In Sec. \ref{obs-bk}, a similar task is performed for  $B \to K  \ell \ell$ decays. Conclusions are provided in Sec. \ref{conc-gen}.

\section{Formalism}
\label{for-tmu}
In this section, we explore the effective Hamiltonian governing $b \to s \ell \ell$ decays, assuming that the WCs associated with new physics exhibit both universal and non-universal interactions with leptons. Additionally, we examine constraints on these WCs derived from current observations within the $b \to s \ell \ell$ sector.

\subsection{Effective Hamiltonian}
 
In the SM, the effective Hamiltonian governing the $b\to s \ell^+ \ell^-$ transition can be expressed as follows:
\begin{eqnarray}
\mathcal{H}^{\rm SM}_{\rm eff} &=& - \frac{\alpha_{em} G_F}{\sqrt{2} \pi} V_{ts}^* V_{tb} \times \nonumber \\
& & \Big[ 2 \frac{C_7^{\rm eff}}{q^2}
 [\overline{s} \sigma^{\mu \nu} q_\nu (m_s P_L  + m_b P_R)b ] \bar{\ell} \gamma_\mu \ell\nonumber \\
& & + C_9^{\rm eff} (\overline{s} \gamma^{\mu} P_L b)(\overline{\ell} \gamma_{\mu} \ell) \nonumber\\ && + C_{10} (\overline{s} \gamma^{\mu} P_L b)(\overline{\ell} \gamma_{\mu} \gamma_5 \ell) \Big] + h.c. \,.
\end{eqnarray}
Here $\alpha_{em}$ is the fine-structure constant, $G_F$ is the Fermi constant, and $V_{ts}$ and $V_{tb}$ are the Cabibbo-Kobayashi-Maskawa (CKM) matrix elements. The $P_{L,R} = (1 \mp \gamma_{5})/2$ are the chiral projection operators and $q$ is the momentum of the off-shell photon in the $b \to s \gamma^* \to s \ell^+ \ell^-$ transition.

If we consider the existence of new physics characterized by vector and axial-vector operators, the effective Hamiltonian governing the $b \to s \ell^+ \ell^-$ process can be written as:
\begin{eqnarray}
\mathcal{H}^{\rm NP}_{\rm eff} &=& -\frac{\alpha_{\rm em} G_F}{\sqrt{2} \pi} V_{ts}^* V_{tb} \left[ C_{9\ell} (\overline{s} \gamma^{\mu} P_L b)(\overline{\ell} \gamma_{\mu} \ell) \right. \nonumber \\
& & \left. + C_{10\ell} (\overline{s} \gamma^{\mu} P_L b)(\overline{\ell} \gamma_{\mu} \gamma_5 \ell)  + C^{\prime}_{9\ell} (\overline{s} \gamma^{\mu} P_R b)(\overline{\ell} \gamma_{\mu} \ell)\right. \nonumber \\
& & \left. + C^{\prime}_{10\ell} (\overline{s} \gamma^{\mu} P_R b)(\overline{\ell} \gamma_{\mu} \gamma_5 \ell)\right] + h.c.  \,\,.
\label{HNP}
\end{eqnarray} 
Here $C_{(9,10)\ell}$ and $C^{\prime}_{(9,10)\ell}$ denote the new physics WCs. With the assumption of the presence of LFU as well as LFUV new physics, the WCs can be expressed as:
\begin{eqnarray}
C_{(9,10)e}&=&C_{(9,10)\tau}=C_{(9,10)}^U\,,\nonumber\\
C_{(9,10)e}^{\prime}&=&C_{(9,10)\tau}^{\prime}=C_{(9,10)}^{\prime U}\,,\nonumber\\
C_{(9,10)\mu}&=& C_{(9,10)}^U +C_{(9,10)\mu}^V\,,\nonumber\\
C_{(9,10)\mu}^{\prime}&=& C_{(9,10)}^{\prime U} +C_{(9,10)\mu}^{\prime V}\,. 
\label{def-wc}
\end{eqnarray}
Thus, the WCs $C^U$ and $C^{\prime U}$ contribute equally to all decays induced by the $b \to s \ell^+ \ell^-$ transitions, while $C^V$ and $C^{\prime V}$ specifically contribute to $b \to s \mu^+ \mu^-$ transition. 
For scenarios with only universal couplings, $C_{(9,10)\mu}^V=C_{(9,10)\mu}^{\prime V}=0$. 

\subsection{Constraints on new physics WCs}
The constraints on these new physics WCs can be derived by conducting a comprehensive global fit to all available $b \to s \ell^+ \ell^-$ measurements. The 1$\sigma$ range of new physics WCs for scenarios which provide a good fit to the current $b \to s \ell \ell$ data are listed in Table \ref{scenarios-both}. We explore two categories of new physics scenarios: one involving only universal couplings to leptons (SU scenarios) and another that includes both universal and non-universal components, with the non-universal components coupling exclusively to muons (S scenarios).

{\rowcolors{2}{white!15!white!30}{yellow!20!white!30}
\begin{table}[hbt]
\addtolength{\tabcolsep}{-1pt}
\begin{center}
\begin{tabular}{|c|c|c|c|}
  \hline\hline
  \rowcolor{lightgray}
Solutions	 & WCs & 1$\sigma$ range & pull   \\
  \hline
SU-I  & $C^{U}_9$ & [-1.19, -0.79] &  4.3 \\ 
\hline
SU-II & $C^{U}_9 = - C^{U}_{10}$ & [-0.51, -0.27]  & 3.3 \\ 
\hline
 SU-III  & $C^{U}_9 = - C^{'U}_{9}$ &[-0.95, -0.60] & 4.1 \\
 \hline
  \hline
S-V  & $C^{V}_{9\mu}$ & [ -1.14, -0.39] &   \\ 
     & $C^{V}_{10\mu}$ & [-0.70, -0.01] & 3.3 \\
     & $C^{U}_9 = C^{U}_{10}$ &  [-0.08, 0.60]  &  \\
\hline
S-VI & $C^{V}_{9\mu} = - C^{V}_{10\mu}$ & [-0.21,-0.1] &  \\ 
     & $C^{U}_9 = C^{U}_{10}$ &[-0.34,-0.05] & 2.8 \\
 \hline
S-VII  & $C^{V}_{9\mu}$ & [-0.32, -0.03]  & \\ 
     & $C^{U}_9$ &[-1.08, -0.6] & 4.5  \\
\hline
S-VIII  & $C^{V}_{9\mu} = - C^{V}_{10\mu}$  &[-0.13,-0.02] &  \\ 
     & $C^{U}_9$ &[-1.13, -0.70]   &4.5 \\
\hline
\hline
S-IX  & $C^{V}_{9\mu} = - C^{V}_{10\mu}$ &[-0.17, -0.05] & \\ 
     & $C^{U}_{10}$  &[-0.03,0.30]    &2.6 \\
\hline
S-X  & $C^{V}_{9\mu}$ &[-0.46, -0.21] & \\ 
     & $C^{U}_{10}$   &[0.04, 0.34]   & 3.3 \\
\hline
S-XI  & $C^{V}_{9\mu}$ & [-0.51, -0.25] & \\ 
     & $C^{\prime U}_{10}$  & [-0.22, 0.02]    & 3.1 \\
\hline
S-XIII  & $C^{V}_{9\mu}$ & [-0.59, -0.27] & \\ 
      & $C^{\prime V}_{9\mu}$ &[0.34,0.02] & \\ 
   & $C^{ U}_{10}$ &[0.02,0.40] & \\ 
   & $C^{\prime U}_{10}$ &[-0.11, 0.23] &3.5 \\ 
 \hline
\end{tabular}
\caption{Allowed new physics solutions assuming new physics couplings to be universal \cite{SinghChundawat:2022zdf} as well as having both universal as well as non-universal components \cite{SinghChundawat:2022ldm}. The pull is defined as $\sqrt{\chi^2_{\rm SM}-\chi^2_{\rm bf}}$, where $\chi^2_{\rm bf}$ represents the $\chi^2$ at the best-fit value in the presence of new physics, and $\chi^2_{\rm SM}$ is the value of $\chi^2$ in the SM. The value of $\chi^2_{\rm SM}$ is approximately 184.}
\label{scenarios-both}
\end{center}
\end{table}

The constraints are obtained from \cite{SinghChundawat:2022zdf} for the framework with only universal couplings and from \cite{SinghChundawat:2022ldm}} for the framework where both universal and non-universal couplings are present. The authors performed a global $\chi^2$-fit to 179 observables in $b \to s \mu^+ \mu^-$ and $b \to s e^+ e^-$ sectors using CERN minimization code {\tt MINUIT} \cite{James:1975dr}. The updated measurements of $R_K$ and $R_{K^*}$ by the LHCb Collaboration in December 2022 \cite{LHCb:2022qnv,LHCb:2022zom} were integrated into the fit, alongside the modified world average of the branching ratio of $B_s \to \mu^+ \mu^-$ \cite{Ciuchini:2022wbq} following the latest measurement from the CMS Collaboration \cite{CMS:2022dbz}. The theoretical predictions of the observables utilized in the fitting process were computed utilizing \texttt{flavio} \cite{Straub:2018kue}, where these observables are pre-implemented based on refs \cite{Bharucha:2015bzk,Gubernari:2018wyi}.  For other global analyses  incorporating the updated LHCb measurement of $R_K$ and $R_{K^*}$, see for e.g., \cite{Ciuchini:2022wbq,Greljo:2022jac,Alguero:2023jeh,Wen:2023pfq,Allanach:2023uxz,Li:2023mrw,Hurth:2023jwr,Das:2023kch,DAlise:2024qmp,Bordone:2024hui,Guadagnoli:2023ddc}

\section{$B \to K^* \ell^+ \ell^-$ observables}
\label{obs-bks}

 The angular distribution of $\overline{B^0} \to \overline{K^{*0}}(\to K^-\pi^+)\ell^+\ell^-$ decay is completely described by four independent observables. These are traditionally chosen to be the three angles ($\theta_{K}$, $\theta_{\ell}$ and $\phi$, as defined in ~\cite{LHCb:2013zuf}) and the invariant mass squared of the dilepton system ($q^2 = (p_B-p_{K^*})^2$). In the notation of ref.~\cite{LHCb:2013zuf}, the full angular decay distribution of $\overline{B^0} \to \overline{K^{*0}}(\to K^-\pi^+)\ell^+\ell^-$ decay is given by
 \begin{equation}
\frac{d^4\Gamma}{dq^2d\cos\theta_{\ell}d\cos\theta_{K}d\phi} = \frac{9}{32\pi}I(q^2,\theta_{\ell},\theta_{K},\phi),
\end{equation}
where
\begin{eqnarray}
I(q^2,\theta_{\ell},\theta_{K},\phi) &=& I^s_1\sin^2\theta_{K} + I^c_1\cos^2\theta_{K}+ \nonumber\\
& &(I^s_2\sin^2\theta_{K}+I^c_2\cos^2\theta_{K})\cos 2\theta_{\ell} \nonumber\\
& & +I_3\sin^2\theta_{K}\sin^2\theta_{\ell}\cos 2\phi \nonumber\\
& &+I_4\sin 2\theta_{K}\sin 2\theta_{\ell}\cos\phi \nonumber\\
& & + I_5 \sin 2\theta_{K}\sin\theta_{\ell}\cos\phi \nonumber \\
& & + I^s_6\sin^2\theta_{K} \cos\theta_{\ell} \nonumber \\
& &+ I_7\sin 2\theta_{K}\sin\theta_{\ell}\sin\phi \nonumber\\
& & + I_8\sin 2\theta_{K}\sin 2 \theta_{\ell} \sin\phi \nonumber \\
& &+I_9\sin^2\theta_{K}\sin^2\theta_{\ell}\sin 2\phi .
\label{Ifunc}
\end{eqnarray}
The twelve $q^2$-dependent angular coefficients $I^{(a)}_i$ \cite{Kruger:1999xa,Altmannshofer:2008dz,Gratrex:2015hna,Ciuchini:2015qxb}
are bilinear combinations of the $K^{*0}$ decay amplitudes, which in turn are functions of WCs and the form factors that depend on the long-distance effects. The functional dependence of the angular coefficients $I^{(a)}_i$ from transversity amplitudes $A$ are defined as~\cite{Altmannshofer:2008dz}:
\begin{eqnarray}
I_1^s &=& \frac{(2+\beta^2_{\ell})}{4}\left[|A^L_{\perp}|^2+|A^L_{\parallel}|^2 +(L\to R)\right] \nonumber \\
&&+ \frac{4m^2_{\ell}}{q^2} {\rm Re}\left(A^L_{\perp}A^{R*}_{\perp}+A^L_{\parallel}A^{R*}_{\parallel}\right), \nonumber 
\end{eqnarray}
\begin{eqnarray}
I^c_1 & = & |A^L_{0}|^2+|A^R_{0}|^2 \nonumber \\
&&+\frac{4m^2_{\ell}}{q^2}\left[|A_t|^2 + 2 {\rm Re}\left(A^L_0 A^{R*}_0\right)\right] ,\nonumber 
\end{eqnarray}
\begin{eqnarray}
I_2^s &=& \frac{\beta^2_{\ell}}{4}\left[|A^L_{\perp}|^2+|A^L_{\parallel}|^2 + (L\to R)\right],\nonumber
\end{eqnarray}
\begin{eqnarray}
I^c_2 &= & -\beta^2_{\ell} \left[|A^L_0|^2 + |A^R_0|^2\right], \nonumber 
\end{eqnarray}
\begin{eqnarray}
I_3 &= & \frac{\beta^2_{\ell}}{2} \left[|A^L_{\perp}|^2 - |A^L_{\parallel}|^2 + (L\to R)\right], \nonumber
\end{eqnarray}
\begin{eqnarray}
I_4 &  = & \frac{\beta^2_{\ell}}{\sqrt{2}} \left[ {\rm Re}(A^L_0 A^{L*}_{\parallel}) + (L\to R)\right],\nonumber
\end{eqnarray}
\begin{eqnarray}
I_5 &= & \sqrt{2} \beta_{\ell} \left[ {\rm Re}(A^L_0 A^{L*}_{\perp}) -(L\to R)\right],\nonumber
\end{eqnarray}
\begin{eqnarray}
I^s_6 &= & 2\beta_{\ell}\left[{\rm Re}(A^L_{\parallel}A^{L*}_{\perp})- (L\to R)\right],\nonumber
\end{eqnarray}
\begin{eqnarray}
I_7 & =& \sqrt{2}\beta_{\ell} \left[{\rm Im}(A^L_0 A^{L*}_{\parallel})- (L\to R)\right],\nonumber
\end{eqnarray}
\begin{eqnarray}
I_8 &= & \frac{\beta^2_{\ell}}{\sqrt{2}}\left[{\rm Im}(A^L_0A^{L*}_{\perp}) + (L\to R)\right],\nonumber
\end{eqnarray}
\begin{eqnarray}
I_9 &= & \beta^2_{\ell}\left[{\rm Im}(A^{L*}_{\parallel}A^L_{\perp})+ (L\to R)\right],
\label{angular-coeff}
\end{eqnarray}
where $\beta_{\ell}= \sqrt{1-4m^2_{\ell}/q^2}$. The expression of transversity amplitudes in terms of the form factors  $V(q^2)$, $A_{0,1,2}(q^2)$ and $T_{1,2,3}(q^2)$ can be found in ref.~\cite{Bharucha:2015bzk}.

The full angular distribution of the $CP$-conjugated mode is given by $B^0 \to K^{*0}(\to K^+\pi^-)\ell^+\ell^-$
 \begin{equation}
\frac{d^4\bar{\Gamma}}{dq^2d\cos\theta_{\ell}d\cos\theta_{K}d\phi} = \frac{9}{32\pi}\bar{I}(q^2,\theta_{\ell},\theta_{K},\phi)\,.
\end{equation}
where $\bar{I}^{(a)}_i$ are the complex conjugate of $I^{(a)}_i$. The relation between $\bar{I}^{(a)}_i$ and $I^{(a)}_i$ can be obtained from the definition of the angles used to describe the decay.
For $\overline{B^0} \to \overline{K^{*0}}(\to K^-\pi^+)\ell^+\ell^-$ decay,  $\theta_{K}$ is the angle between the directions of kaon in the $\overline{K^{*0}}$ rest frame and the  $\overline{K^{*0}}$ in the rest frame of $\overline{B}$.  The angle $\theta_{\ell}$ is between the directions of the $\ell^-$ in the dilepton rest frame and the dilepton in the rest frame of $\overline{B}$ whereas the angle $\phi$ is the azimuthal angle between the plane containing the dilepton pair and the plane encompassing the kaon and pion from the $\overline{K^{*0}}$. For $B^0 \to K^{*0}(\to K^+\pi^-)\ell^+\ell^-$ decay mode,  $\theta_{K}$ is the angle between the directions of kaon in the ${K^{*0}}$ rest frame and the  ${K^{*0}}$ in the rest frame of ${B}$ whereas the angle $\theta_{\ell}$ is between the directions of the $\ell^+$ in the dilepton rest frame and the dilepton in the rest frame of ${B}$. This leads to the following transformation of angular coefficients under $CP$ \cite{Belle-II:2018jsg}
\begin{equation}
I^{(a)}_{1,2,3,4,5,6} \Longrightarrow \bar{I}^{(a)}_{1,2,3,4,5,6}, \quad I^{(a)}_{7,8,9} \Longrightarrow -\bar{I}^{(a)}_{7,8,9},
\end{equation}
 Combining $B^0$ and $\overline{B}^0$ decays, one can construct the CP-averaged angular observables \cite{Altmannshofer:2008dz}
\begin{equation}
S^{(a)}_i = \frac{I^{(a)}_i(q^2)+ \bar{I}^{(a)}_i(q^2)}{d(\Gamma +\bar{\Gamma})/dq^2}.
\label{sdef}
\end{equation}
The difference of these angular coefficients will result in corresponding $CP$-violating angular observables\cite{Bobeth:2008ij,Altmannshofer:2008dz}. 

Several well-established observables in the decay of $B \to K^* \ell^+ \ell^-$ can be expressed in terms of angular coefficients 
$I^{(a)}_i$ as well as $CP$-averaged angular observables $S^{(a)}$:
\begin{itemize}
\item The angular-integrated differential decay rate can be written as
\begin{eqnarray}
\frac{d\Gamma}{dq^2}&=&\int d\cos \theta_{\ell}\, d\cos \theta_K \,d\phi \frac{d^4\Gamma}{dq^2\, d\cos \theta_K \,d\cos \theta_{\ell}\, d\phi}\nonumber\\ &=&\frac{3}{4}\left(2I^s_1+I^c_1\right) - \frac{1}{4}\left(2I_2^s+I_2^c\right)\,.
\end{eqnarray}

\item The normalized forward-backward asymmetry can be expressed as
\begin{eqnarray}
A_{FB}&=& \left[\int_0^1-\int_{-1}^0\right] d\cos \theta_{\ell} \frac{d^2(\Gamma -\bar{\Gamma})}{dq^2d\cos \theta_{\ell}}/\frac{d(\Gamma +\bar{\Gamma})}{dq^2}
\nonumber\\ &=&\frac{3}{4} S^s_6\,.
\label{afb-tau}
\end{eqnarray}

\item The $K^*$ longitudinal polarization fraction can be written as
\begin{eqnarray}
f_L = \frac{3S^c_1 - S^c_2}{4}\,.
\label{fl-tau}
\end{eqnarray}
\end{itemize}
\begin{figure*} [htb]
 \centering
\includegraphics[scale=.6]{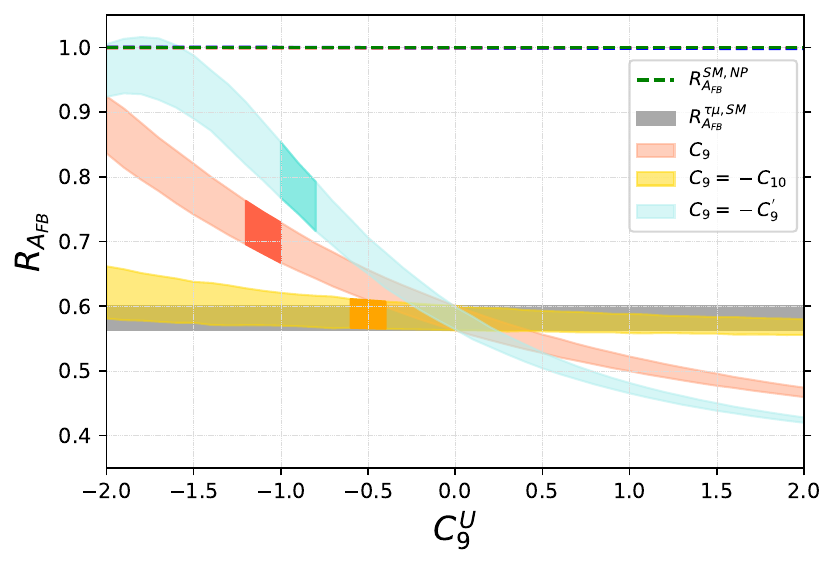}
\includegraphics[scale = 0.6]{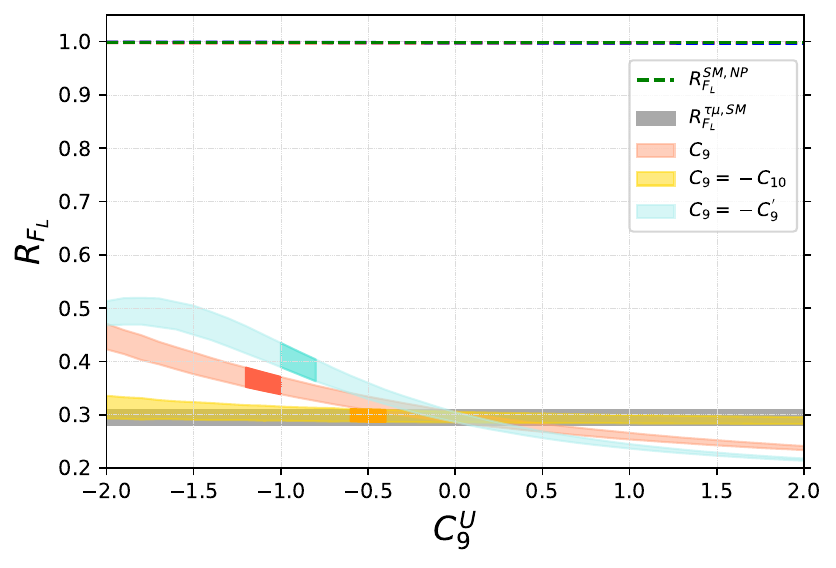}
    \caption{ The left and right panels respectively demonstrate the dependence of the ratios $R_{A_{FB}}^{\tau \mu}$ and $R_{f_L}^{\tau \mu}$ on new physics WCs with purely universal components. The grey-shaded regions represent the predictions within the SM framework. In contrast, the lighter bands in red, yellow, and blue colours are associated with the new physics scenarios characterized by $C_9^U$, $C_9^U = - C_{10}^U$, and $C_9^U = - C_{9}^{'U}$, respectively.  These bands encapsulate the theoretical uncertainties inherent in these calculations. The darker shaded regions in red, yellow, and blue indicate the 1$\sigma$ intervals of respective NP coupling range allowed by current experimental measurements in the $b \to s \ell \ell$ processes ($\ell=e,\,\mu$). For comparative purposes, the plots also include the predicted values for $R_{A_{FB}}^{\mu e}$ and $R_{f_L}^{\mu e}$ ratios under universal coupling scenarios, demonstrating their consistency with the SM expectations. }
    \label{kstau-plot1}
\end{figure*}

The $S^{(a)}$ observables are sensitive to hadronic uncertainties mainly due to the form factors \cite{Khodjamirian:2010vf,Bharucha:2015bzk,Gubernari:2018wyi} and nonlocal contributions associated with charm-quark loops \cite{Beneke:2001at,Khodjamirian:2010vf,Descotes-Genon:2014uoa,Ciuchini:2015qxb,Capdevila:2017ert,Bobeth:2017vxj,Blake:2017fyh,Gubernari:2020eft,Ciuchini:2022wbq,Ciuchini:2021smi,Gubernari:2022hxn}.  The calculations in refs. \cite{Gubernari:2020eft,Gubernari:2022hxn} which build upon \cite{Khodjamirian:2010vf,Bobeth:2017vxj},  primarily address the ``charm-loop"-to-\(\gamma^*\) (\(q^2\)) amplitude, with long-distance effects manifesting as poles and cuts in the \(q^2\) variable. On the other hand, refs. \cite{Ciuchini:2022wbq,Ciuchini:2015qxb,Ciuchini:2021smi} emphasize  the significance of contributions from \(B \rightarrow\) di-meson rescatterings, corresponding to cuts in the full decay variable \((q + k)^2\), where \(k\) represents the momentum of the final-state \(K^{(*)}\) \cite{Guadagnoli:2023ddc}. In the current analysis, we focus on the magnitude of the long-distance contributions stemming from rescattering of intermediate states.

The form factors in the low-$q^2$ region are calculated using light-cone sum rules (LCSR) or light-meson distribution amplitudes whereas in the high-$q^2$ region, form factors are obtained from lattice computations \cite{Horgan:2013hoa,Flynn:2015ynk}. One can construct optimized observables with reduced uncertainties by proper combination of $f_L$ and $S^{(a)}$. These observables have been proposed by several groups, see for e.g., \cite{Kruger:2005ep,Egede:2008uy,Bobeth:2011gi,Becirevic:2011bp,Matias:2012xw,Descotes-Genon:2013vna,Descotes-Genon:2013wba}. A frequently used form is given in \cite{Descotes-Genon:2013vna,Descotes-Genon:2013wba}. A generalized and extensive analysis of angular distribution formalism can be found in ref. \cite{Gratrex:2015hna}. In this work, for $B \to K^* \tau^+ \tau^-$ decay, we consider the following set of optimized observables $P^{(')}_i$  defined in ref. \cite{Descotes-Genon:2013vna,Descotes-Genon:2013wba} and written in the basis of \cite{Kstarlhcb2}:
\begin{eqnarray}
P_1 &=& \frac{S_3}{2 S_2^s}, \,\, P_2=\frac{S_6^s}{8 S_2^s}, \,\, P_3=\frac{S_{9}}{4 S_2^s}\,,\nonumber\\
P'_4  &=& \frac{S_4}{2\sqrt{-S_2^s S_2^c}}, \,\,P'_5  = \frac{S_5}{2\sqrt{-S_2^s S_2^c}}\,,\nonumber\\
P'_6  &=& \frac{S_7}{2\sqrt{-S_2^s S_2^c}}, \,\, P'_8 = \frac{S_8}{2\sqrt{-S_2^s S_2^c}}\,.
\label{def:opt}
\end{eqnarray}
 The theoretical predictions of the observables are computed utilizing \texttt{flavio} \cite{Straub:2018kue}, where these observables are pre-implemented based on refs \cite{Bharucha:2015bzk,Gubernari:2018wyi}.

We will now consider the LFUV ratios of the above observables in the $\tau-\mu$ sector and determine whether they can be considered true LFUV observables. We first consider the ratio of partial widths:
\begin{equation}
    R_{K^*}^{\tau \mu} =  \frac{\Gamma(B \to K^* \tau^+ \tau^-)}{\Gamma(B \to K^*  \mu^+ \mu^-)} \,.
\end{equation}

It has been demonstrated that this observable should not be considered as a true LFUV observable because it can deviate from its SM prediction even for new physics scenarios with only universal lepton couplings~\cite{Alok:2023yzg}.  
This deviation is attributed to the fact that the various terms in the expression for the decay rates exhibit different dependencies on the lepton mass. This is referred to as ``mass effects''. 

We now analyze additional ratios to determine whether they are genuine LFUV observables. Specifically, we examine the ratio of forward-backward asymmetries $R_{A_{FB}}^{\tau \mu}$ defined as
\begin{equation}
    R_{A_{FB}}^{\tau \mu} \equiv \frac{\langle A_{FB} \rangle(B \to K^* \tau^+ \tau^- )}{\langle A_{FB} \rangle(B \to K^* \mu^+ \mu^-)}\,.
\end{equation}

Here, we consider the same integration interval for $\langle A_{FB} \rangle$ in both the numerator and denominator which is [15-19] $\text{GeV}^2$. The same approach is followed for all $B \to K^* \ell^+ \ell^-$ LFUV ratios examined in this work.
As evident from eqs. \eqref{afb-tau} and \eqref{sdef}, the numerator of $A_{FB}$ depends upon the  angular coefficient $I_6^s$ whereas the denominator depends upon $\Gamma(B \to K^* \ell \ell)$ which is a linear combination of coefficients of $I_1^s$, $I_1^c$, $I_2^s$ and $I_2^c$. The angular coefficient $I_6^s$ does not have an explicit dependence on the lepton mass $m_{\ell}$ apart from having a common multiplicative factor $\beta_\ell$ which is the same for all WCs. However, in the denominator,  different terms have distinct dependence on $m_\ell$. Therefore it becomes apparent that similar to $R_{K^*}^{\tau \mu}$, the  ratio $R_{A_{FB}}^{\tau \mu}$  
exhibits dependence on lepton masses i.e., suffers from mass effects. Consequently, it is necessary to assess numerically whether this observable qualifies as a genuine LFUV observable. To facilitate this evaluation, we analyze the predictions of $R_{A_{FB}}^{\tau \mu}$ for the selected new physics scenarios.

 These follow from a data-driven approach where we consider those 1D scenarios that describe the current $b \to s \ell \ell$ measurements better than the SM. These scenarios may change with updated experimental measurements and/or theoretical predictions; however, if a particular LFUV observable deviates from the SM prediction even for just one single NP  scenario with universal couplings, the observable cannot be considered as a good probe of LFU violation. 

 It is evident from the Fig. \ref{kstau-plot1} that $R_{A_{FB}}^{\tau \mu}$ deviates from its SM prediction even for the universal couplings. The deviation increases with increasing values of these new physics couplings and is more prominent for the $C_9^U$ and $C_9^U=-C_9^{'U}$ scenarios as compared to the $C_9^U=-C_{10}^{U}$ scenario.  Further, already within the currently allowed range of new physics couplings for various scenarios (as given in Table \ref{scenarios-both} and presented in Fig.~\ref{kstau-plot1} by darker shaded regions)  $R_{A_{FB}}^{\tau \mu}$ is deviating from its SM prediction. Therefore mere deviation of $R_{A_{FB}}^{\tau \mu}$ from the SM cannot be attributed to LFUV type of new physics. On the other hand, it is also evident from the figure that $R_{A_{FB}}^{\mu e}$  doesn't deviate from its SM prediction, indicating that it is a genuine LFUV observable in the $\mu-e$ sector.

\begin{figure*} [htb]
 \centering
\includegraphics[scale=.5]{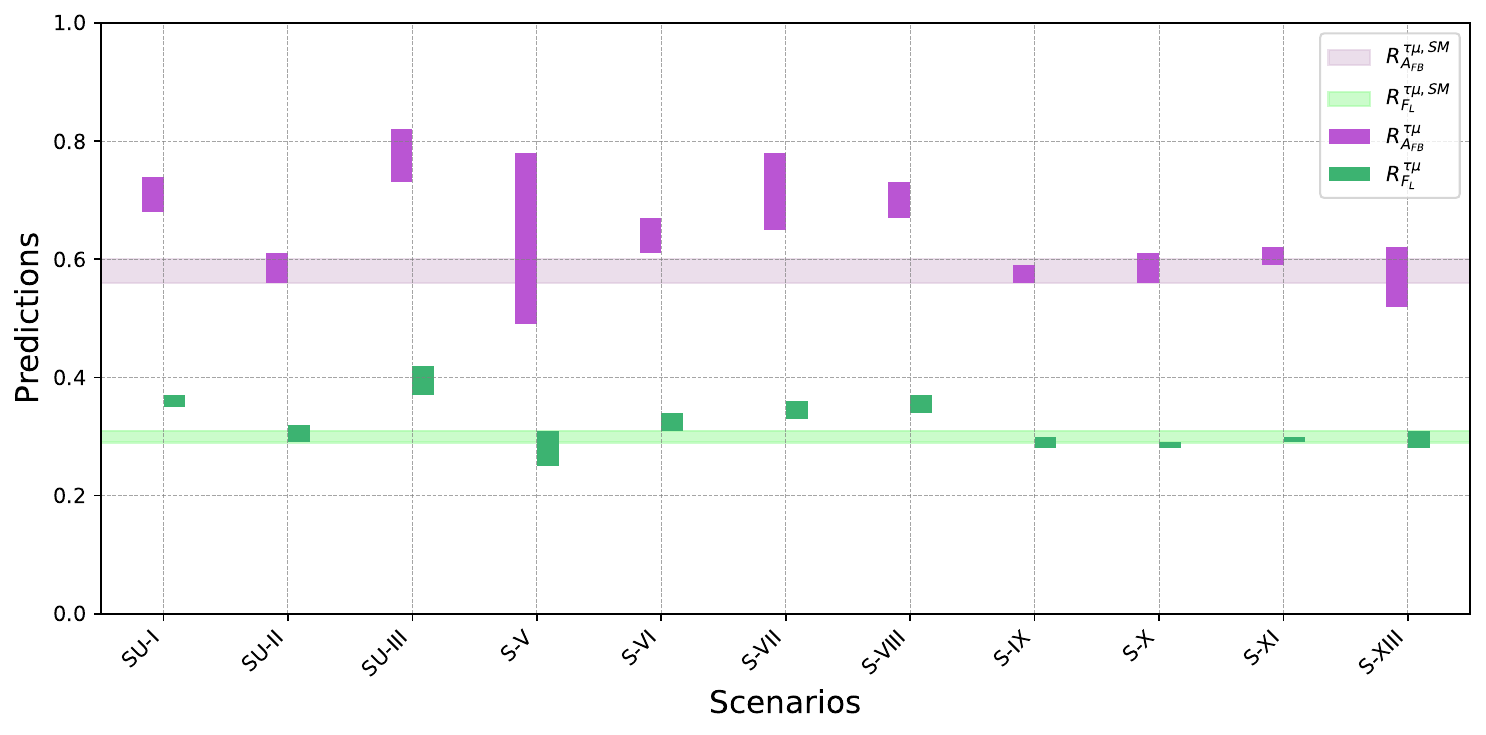}
    \caption{Predicted ranges (1$\sigma$) for $R_{A_{FB}}^{\tau \mu}$ and $R_{f_L}^{\tau \mu}$ across all viable solutions, considering frameworks with only universal couplings and those encompassing both universal and non-universal couplings. }
    \label{kstau-plot2}
\end{figure*}

Hence, to harness the discriminatory power of $R_{A_{FB}}^{\tau \mu}$ in discerning between universal and non-universal types of new physics, it is imperative to scrutinize the projections of $R_{A_{FB}}^{\tau \mu}$ across all data-driven favored new physics scenarios pertaining to both categories. Only through this process can one effectively differentiate between these two classes of new physics. The depiction of this scenario for current $b \to s \ell \ell$ measurements is illustrated in Fig.~\ref{kstau-plot2}. 

It is evident from the figure that all new physics scenarios with only universal couplings,  favored by the current data, predict $R_{A_{FB}}^{\tau \mu} \gtrsim R_{A_{FB}}^{\tau \mu, \, {\rm SM}}$. The SU-I and SU-III scenarios, as listed in Table \ref{scenarios-both},  predict $R_{A_{FB}}^{\tau \mu} > R_{A_{FB}}^{\tau \mu, \, {\rm SM}}$ whereas for SU-II scenario, $R_{A_{FB}}^{\tau \mu} \approx R_{A_{FB}}^{\tau \mu, \, {\rm SM}}$. Thus none of the scenarios favoured by the current $b \to s \ell \ell $ data allows $R_{A_{FB}}^{\tau \mu} < R_{A_{FB}}^{\tau \mu, \, {\rm SM}}$. For the framework where both universal and non-universal components are present, the scenarios favoured by the current data allows for $R_{A_{FB}}^{\tau \mu} > R_{A_{FB}}^{\tau \mu, \, {\rm SM}}$ as well as $R_{A_{FB}}^{\tau \mu} < R_{A_{FB}}^{\tau \mu, \, {\rm SM}}$. For the S-V and S-XIII scenarios, the predicted values of $R_{A_{FB}}^{\tau \mu}$ can be lower than the SM predictions but overlap with the SM values within the error bars. For all other scenarios, the predicted value of $R_{A_{FB}}^{\tau \mu}$ are either consistent or greater than the SM. 
Consequently, if the experimentally measured value of $R_{A_{FB}}^{\tau \mu}$ turns out to be lower than its SM prediction, it would indicate that the new physics responsible for this deviation must involve a non-universal component alongside the universal one. However, if the measured value of  $R_{A_{FB}}^{\tau \mu}$  turns out to be greater than $R_{A_{FB}}^{\tau \mu, \, {\rm SM}}$,  it would be challenging to distinguish between the two classes of solutions.

The next observable under consideration is $R_{f_L}^{\tau \mu}$, which is defined as
\begin{equation}
 R_{f_L}^{\tau \mu} \equiv \frac{{\langle f_L \rangle}(B \to K^* \tau^+ \tau^- )}{\langle f_L \rangle(B \to K^* \mu^+ \mu^-)}\,.
\end{equation}
As evident from eq.\eqref{fl-tau}, $f_L$ depends on the angular coefficients $S^c_2$, which, in turn, depends on $I^c_2$ and $\Gamma (B \to K^* \ell \ell)$. Although $I^c_2$ lacks explicit dependence on lepton mass except for  $\beta_\ell$ factor which is a common multiplicative factor to all WCs appearing in  $I^c_2$, $\Gamma (B \to K^* \ell \ell)$ explicitly depends on lepton mass. Hence, the ratio $R_{f_L}^{\tau \mu}$  may not qualify as a genuine LFUV observable akin to $R_{K^*}^{\tau \mu}$ and $R_{A_{FB}}^{\tau \mu}$.  This assertion is indeed supported by the observations from the right panel of Fig.~\ref{kstau-plot1}. It is evident that $R_{f_L}^{\tau \mu}$ deviates from its SM prediction even for new physics scenarios where WCs only have universal components, i.e., universal couplings to leptons. Like 
$R_{A_{FB}}^{\tau \mu }$, this deviation is more pronounced in SU-I and SU-III scenarios compared to the SU-II scenario. The deviation of $R_{f_{L}}^{\tau \mu}$ from its SM value becomes more prominent for larger values of the universal WCs. On the other hand, the figure indicates that $R_{f_L}^{\mu e}$ does not deviate from its SM prediction, suggesting that it is a genuine LFUV observable in the $\mu-e$ sector.

Again, it is imperative to utilize comparisons of predictions for $R_{f_L}^{\tau \mu}$ across all favored new physics scenarios to distinguish between scenarios involving LFU and those involving LFUV.  In the present context, the predictions of $R_{f_L}^{\tau \mu}$ for all the considered new physics scenarios are illustrated in Fig.~\ref{kstau-plot2}. 

For frameworks with only universal couplings to leptons, the predictions of $R_{f_L}^{\tau \mu}$ suggests that $R_{f_L}^{\tau \mu} \gtrsim R_{f_L}^{\tau \mu,\, {\rm SM}}$ for all solutions which provide a better fit to the current $b \to s \ell \ell$ data as compared to the SM. For the framework with both universal and non-universal WCs, the predicted values of $R_{f_L}^{\tau \mu}$ can be larger than the SM value for S-VI, S-VII and S-VIII scenarios whereas the solutions S-IX, S-X, S-XI, and S-XIII predict $R_{f_L}^{\tau \mu}$ to be close to the SM. In the case of the new physics scenario S-V, the predicted values of $R_{f_L}^{\tau \mu}$ can be lower than that of the SM, although they still align with the corresponding SM values within the error bar. Thus, these two frameworks can be distinguished if the experimentally measured $R_{f_L}^{\tau \mu}$ is found to be below the SM value. Should $R_{f_L}^{\tau \mu}$ exceed the SM value, discrimination between the frameworks would not be feasible.

It is evident that neither $R_{A_{FB}}^{\tau \mu}$ nor $R_{f_L}^{\tau \mu}$ qualify as genuine LFUV observables. This is primarily because both of these observables rely on $\Gamma (B \to K^* \ell \ell)$, and as previously discussed, $\Gamma (B \to K^* \ell \ell)$ is contingent on angular coefficients, wherein different terms exhibit distinct dependencies on lepton mass. Consequently, any $\tau-\mu$ LFU ratios featuring $\Gamma (B \to K^* \ell \ell)$  may not be deemed as genuine LFU testing observables.

It is important to note that when defining these LFUV ratios, the integration limits for the numerator and denominator are taken to be the same. In this context, ``mass effect" refers to the presence of the lepton mass term in various coefficients, leading to differences between the $\tau$ and $\mu$ channel observables. However, if we redefine these LFUV observables by using different integration ranges for the numerators and denominators, for example, by considering the full kinematic region $4m_{\ell}^2 \leq q^2 \leq (m_B - m_{K^*})^2$, an additional ``mass effect" arises due to the significant difference in the integration range of the numerator and denominator\footnote{ Within certain \( q^2 \) regions,  the shape of the differential decay width near the kinematic endpoint is sensitive to $m_\ell$. For example, in the prediction for $R^{\mu e}_{K^*}$ in the \( 0.045 \, \text{GeV}^2 \leq q^2 \leq 1.1 \, \text{GeV}^2 \) bin, the kinematic threshold of the muon mode and the rapid variation of \( {d\Gamma}/{dq^2} \) close to this threshold result in larger theoretical uncertainties \cite{Bordone:2016gaq}.}Therefore, if we define all the LFUV ratios considering the full kinematic range, the mass effect will exhibit a ``two-fold" nature.

Next, we examine the ratios of the optimized observables, denoted as $R_{P_i}^{\tau \mu}$. Since all optimized observables are constructed as ratios of angular coefficients $S_i$, there exists no explicit dependence on $\Gamma(B \to K^* \ell \ell)$. Hence, it becomes necessary to scrutinize each of these ratios individually to determine whether they qualify as genuine LFUV observables.

The optimized observable $P_1$ is the ratio of coefficients $I_3$ and $I_2^s$. Examining eq. \eqref{angular-coeff}, it's apparent that both $I_3$ and $I_2^s$ involve lepton masses only within $\beta_\ell$, which serves as the common multiplicative factor to the amplitudes $A^{L,R}$. Consequently, the ratio $R_{P_1}^{\tau \mu} \equiv \langle P^\tau_1 \rangle / \langle P^\mu_1 \rangle$ remains unaffected by mass effects. Therefore, it can be regarded as a genuine LFUV observable, implying that any deviation from its SM expectation can solely stem from LFUV-type new physics.

The observable $P_2$ is the ratio of angular coefficients $I_6^s$ and  $I_2^s$. Apart from the common multiplicative factor $\beta_\ell$, these coefficients do not show any explicit dependence on the lepton mass
 $m_\ell$. Since $P_2$ itself is independent of $m_\ell$, the ratio $R_{P_2}^{\tau \mu} \equiv \langle P^\tau_2 \rangle / \langle P^\mu_2 \rangle$ like
$R_{P_1}^{\tau \mu} $  does not depend on the lepton mass, thereby confirming that it is a genuine LFUV observable. The same is true for the observable $R_{P_3}^{\tau \mu} \equiv \langle P^\tau_3 \rangle/ \langle P^\mu_3 \rangle$ as the angular coefficients $I_9$ and  $I_2^s$ do not exhibit dependence on $m_\ell$, apart from a common factor $\beta_\ell$. This lack of dependence on the lepton mass for these ratios underscores their potential utility in probing new physics through the lens of LFUV new physics.

We now focus on the observable $P'_4$. Here, the numerator depends upon $I_4$, and the denominator depends on the square root of the product of $I_s^2$ and $I_c^2$. In these angular coefficients, there is no dependence on \(m_\ell\) except for \(\beta_\ell\). Hence $P'_4$ is independent of $m_\ell$  implying that $R_{P'_4}^{\tau \mu} \equiv \langle P'^\tau_4 \rangle / \langle P'^\mu_4 \rangle$ is a genuine observable to test LFUV new physics.

The same inference is applicable to other LFU ratios of the optimized observables $P'_5$, $P'_6$, and $P'_8$, which are defined respectively as $R_{P'_5}^{\tau \mu} \equiv \langle P'^\tau_5 \rangle / \langle P'^\mu_5 \rangle$, $R_{P'_6}^{\tau \mu} \equiv \langle P'^\tau_6 \rangle / \langle P'^\mu_6 \rangle$, and $R_{P'_8}^{\tau \mu} \equiv  \langle P'^\tau_8 \rangle / \langle P'^\mu_8 \rangle$. These ratios are inherently free from mass effects, serving as robust indicators of LFUV. This attribute originates from the mathematical structure of their numerators and denominators. The denominator involves the term $\sqrt{I_s^2 I_c^2}$, which does not depend on the lepton mass $m_\ell$, except through the common kinematic factor $\beta_\ell$. Furthermore, in the numerators of $P'_5$, $P'_6$, and $P'_8$, the lepton mass $m_\ell$  appears only through $\beta_\ell$. Consequently, the presence of $m_\ell$ in these ratios is effectively neutralized, ensuring that the expressions for $R_{P'_5}^{\tau \mu}$, $R_{P'_6}^{\tau \mu}$, and $R_{P'_8}^{\tau \mu}$ are free from mass effect.

Thus, the LFUV ratios of all optimized observables, as delineated in eq.~\eqref{def:opt}, can be definitively considered as genuine LFUV observables in the $\tau-\mu$ sector. These ratios are meticulously crafted to exclude mass effects, thereby providing a true measure of LFUV. The mathematical structure of these observables ensures that any deviations from the SM predictions can be attributed to genuine differences in lepton interactions rather than artifacts of lepton mass. This positions them as critical tools in the search for new physics within the $\tau-\mu$ interactions, offering insights into potential LFUV and hence enhancing our understanding of the symmetry structure of fundamental interactions beyond the current paradigm.
\section{$B \to K \ell^+ \ell^-$ observables}
\label{obs-bk}
The full angular distribution of $B \to K \ell \ell$ decay can be written as \cite{Becirevic:2012fy,Bobeth:2007dw}
\begin{eqnarray}
    \frac{d^2\Gamma}{dq^2 d \cos \theta} = a_\ell(q^2) + b_\ell(q^2)  \cos \theta+ c_\ell (q^2) \cos^2\theta,
\end{eqnarray}
where $q^2=(p_{\ell^+}+p_{\ell^-})^2$ and $\theta$ is angle between the direction of $\bar{B}$ and of $\ell^-$ in the center of mass frame of the lepton. The coefficients, $a_\ell(q^2), \, b_\ell(q^2)$ and $ c_\ell (q^2)$ are defined as
\begin{eqnarray}
a_{\ell}(q^2) &=& E(q^2) \Bigg[ q^2 |F_P(q^2)|^2 + \frac{\lambda_K}{4} \left( |F_V(q^2)|^2 + |F_A(q^2)|^2\right)\nonumber\\
&&+ 2m_{\ell} \left( m_B^2 -m_K^2 +q^2\right){\rm Re} (F_P(q^2) F_A^{*}(q^2)) \nonumber\\
&& + 4 m_{\ell}^2 m_B^2 |F_A(q^2)|^2  \Bigg]\,,\\
b_{\ell}(q^2) &=& 0,\\
c_{\ell}(q^2) &=& -\frac{\lambda_K}{4} \beta_{\ell}^2 E(q^2) \left( |F_V(q^2)|^2 + |F_A(q^2)|^2\right)\,,
\end{eqnarray}
where
\begin{equation}
E(q^2) = \frac{G_F^2\alpha^2 |V_{tb} V_{ts}^*|^2}{512 \pi^5 m_B^3}\beta_{\ell}\sqrt{\lambda_K},
\end{equation}
with $\lambda_K = m^4_B+m^4_{K}+q^4-2(m^2_B m^2_{K} +m^2_{B}q^2+m^{2}_{K}q^2)$, $\beta_{\ell}= \sqrt{1-4m^2_{\ell}/q^2}$ and the $F(q^2)$ functions are defined in terms of WCs and form-factors and are given as 
\begin{eqnarray}
F_V(q^2) &=& (C_9^{\rm eff} + C_{9\ell}+C'_{9\ell}) f_+(q^2) \nonumber\\ 
&+& \frac{2 m_b C_7^{\rm eff} }{m_B + m_K}f_T(q^2),\\
F_A(q^2) &=& (C_{10} + C_{10\ell}+C'_{10\ell}) f_+(q^2), \\
F_P(q^2) &=& - m_\ell (C_{10} + C_{10\ell}+C'_{10\ell}) \nonumber\\
&\times&  \left[f_+(q^2) - \frac{m_B^2-m_K^2}{q^2} (f_0(q^2)-f_+(q^2))\right]\,.
\end{eqnarray}
The $B \to K$ form-factors $f_0(q^2)$,  $f_+(q^2)$ and $f_T(q^2)$ are defined  in \cite{{Becirevic:2012fy,Bobeth:2007dw}}. In the low-$q^2$ region, all form-factors reduce to one soft form-factor \cite{Charles:1998dr,Beneke:2000wa}. In the high-$q^2$ region too, symmetry relations among the form factors can be delved with the improved Isgur-Wise relation \cite{Bobeth:2011nj}. Here it should be noted that the function $b_\ell (q^2)$ should be non-zero only in the presence of scalar and tensor couplings \cite{Bobeth:2007dw,Alok:2008wp}. Based on the above angular distribution, we can define the following observables:
\begin{itemize}
    \item The decay rate of $B \to K \ell^+ \ell^-$ which is given by 
\begin{equation}
\Gamma (B \to K \ell^+ \ell^-) = \int_{q^2_{\rm min}}^{{q^2_{max}}} dq^2 \left(2a_{\ell}(q^2) +\frac{2}{3} c_{\ell}(q^2)\right)\,.
\end{equation}

\item The observable $F^\ell_H$ which is defined as
\begin{equation}
F^\ell_H = \frac{\int_{q^2_{\rm min}}^{{q^2_{max}}} dq^2 \left(a_{\ell}(q^2) + c_{\ell}(q^2)\right)}{\int_{q^2_{\rm min}}^{{q^2_{max}}} dq^2 \left(a_{\ell}(q^2) +\frac{1}{3} c_{\ell}(q^2)\right)}\,.
\label{fh-bkll}
\end{equation}

\end{itemize}
The forward-backward asymmetry of leptons, involving the coefficient $b_\ell(q^2)$, is $\approx 0$ in the SM and also with the introduction of new physics in the form of vector and axial-vector operators.

\begin{figure} [htb]
 \centering
\includegraphics[scale = 0.6]{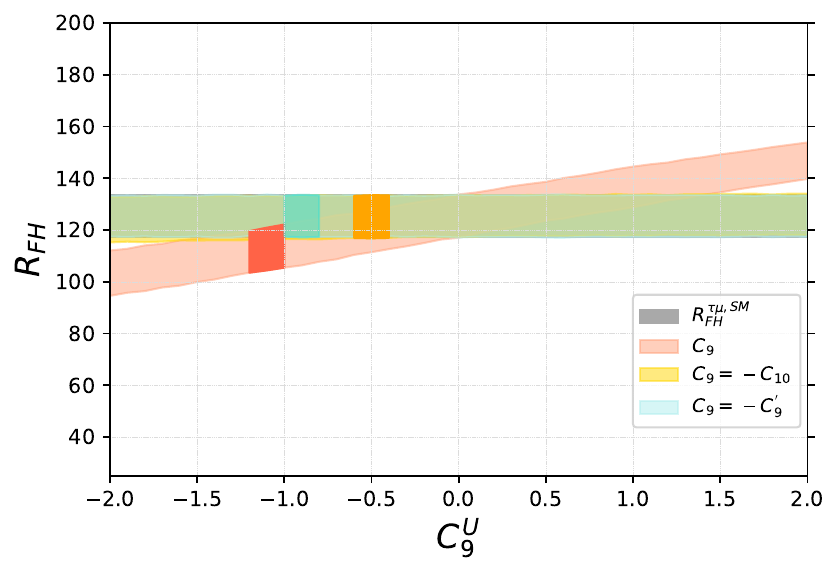}
\caption{The illustration highlights how the ratio \( R_{F_H}^{\tau \mu} \) varies with the new physics WCs that have purely universal components. The grey band represents the SM prediction. In contrast, the lighter shaded regions in red, yellow, and blue correspond to new physics scenarios involving \( C_9^U \), \( C_9^U = -C_{10}^U \), and \( C_9^U = -C_{9}^{'U} \), respectively. The darker shaded regions in the same colors indicate the 1$\sigma$ intervals of the respective new physics coupling ranges permitted by current experimental measurements in the \( b \to s \ell \ell \) processes (\(\ell = e, \mu\)). The predicted values for the \( R_{F_H}^{\mu e} \equiv {F_H^\mu}/{F_H^e} \) ratio are not included because its value becomes very large as \( F_H^\ell \) approaches exceedingly small values when \( m_\ell \to 0 \).
 Nevertheless, it has been verified that its value does not deviate from the SM for universal couplings.}
 \label{fh-plot1}
\end{figure}

We now consider the following LFUV ratio
\begin{equation}
   R_K^{\tau \mu} \equiv \frac{\Gamma(B \to K \tau^+ \tau^-)}{\Gamma(B \to K \mu^+ \mu^-)}\,.
\end{equation}
In \cite{Alok:2023yzg}, it was shown that this observable cannot be termed as a genuine LFUV observable as it deviates from its SM prediction even for new physics scenarios with only universal couplings to leptons. This is because, in the expression of $\Gamma (B \to K \ell^+ \ell^-)$ which is a linear combination of coefficients $a_\ell(q^2)$ and $c_\ell(q^2)$,  apart from $\beta_\ell$ which appears as a common multiplicative factor to all WCs,  different terms have distinct dependence on $m_\ell$.

Therefore, we are now left with the observable $F^\ell_H $ to see whether it can be utilized to construct a LFU ratio observable. We define $R_{F_H}^{\tau \mu}\equiv F^\tau_H/F^\mu_H$. Here, we consider the same integration interval for $F_H^\ell$ in both the numerator and denominator, which is [15-22] $\text{GeV}^2$. The same approach is followed for other $B \to K \ell^+ \ell^-$ LFUV ratios as well. As evident from eq.~\eqref{fh-bkll}, the numerator and denominator of $F^\ell_H $ observable are different linear combinations of coefficients $a_\ell(q^2)$ and $c_\ell(q^2)$. As different terms appearing in these combinations have distinct dependence on $m_\ell$, the LFU ratio constructed using this observable would suffer from mass effects and hence cannot be classified as a genuine LFU ratio.

This is also evident from Fig. \ref{fh-plot1}, which shows that the LFUV ratio \( R_{F_H}^{\tau \mu} \) deviates from its SM prediction even for new physics scenarios involving only universal couplings. The value of \( R_{F_H}^{\tau \mu} \) deviates from the SM across the entire range of the new physics WC \( C_9^U \), except for a narrow range around \( C_9^U \approx 0 \). The deviation increases significantly for larger values of \( C_9^U \). In the case of the \( C_9^U = -C_{10}^U \) solution, \( R_{F_H}^{\tau \mu} \) remains consistent with the SM prediction across the entire range of WCs considered, including the 1$\sigma$ allowed region for \( C_9^U = -C_{10}^U \). Additionally, the \( C_9^U = -C_{9}^{'U} \) scenario predicts \( R_{F_H}^{\tau \mu} \approx R_{F_H}^{\tau \mu, \rm SM} \) throughout the range of examined WCs.

\begin{figure} [htb]
 \centering
\includegraphics[scale = 0.6]{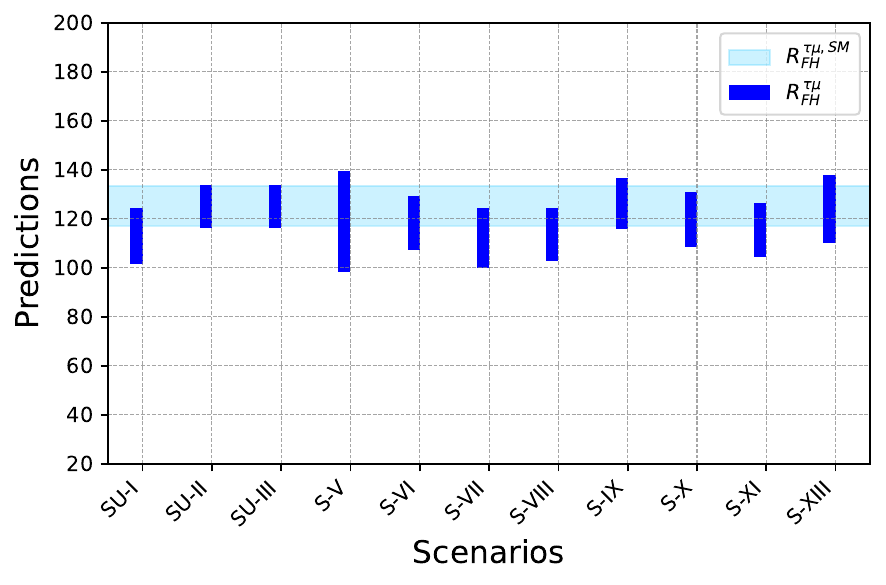}
\caption{The predicted 1$\sigma$ ranges for \( R_{F_H}^{\tau \mu} \) are shown for all feasible solutions, taking into account frameworks that include only universal couplings as well as those that incorporate both universal and non-universal couplings.
}
 \label{fh-plot2}
\end{figure}

 It is therefore crucial to compare the predictions for \( R_{F_H}^{\tau \mu} \) across all preferred new physics scenarios to differentiate between those involving LFU and LFUV. Fig. \ref{fh-plot2} illustrates the predictions for \( R_{F_H}^{\tau \mu} \) for each new physics scenario under consideration.
 For frameworks with only universal lepton couplings, the predictions for \( R_{F_H}^{\tau \mu} \) indicate that \( R_{F_H}^{\tau \mu} \lesssim R_{F_H}^{\tau \mu,\, \rm{SM}} \) for all solutions that provide a better fit to current \( b \to s \ell \ell \) data compared to the SM. The SU-II and SU-III scenarios have predictions within the SM band, showing no significant deviation from the SM expectations. In contrast, the SU-I scenario predicts values that may fall outside the SM band, suggesting potential deviations due to new physics effects.

Scenarios with both universal and non-universal couplings exhibit a wide range of predictions: some lie within the SM band, some show moderate deviations, and others exhibit large deviations. Scenario S-IX aligns with the SM predictions, indicating no significant deviation, while moderate deviations from the SM are possible for scenarios S-VI, S-X, S-XI and S-XIII. The predicted values for the S-V, S-VII and S-VIII scenarios overlap with the SM band but can extend well outside it, indicating a noticeable suppression, \( R_{F_H}^{\tau \mu} < R_{F_H}^{\tau \mu,\, \rm{SM}} \). However, the maximum allowed suppression in \( R_{F_H}^{\tau \mu} \) compared to the SM is almost the same as that allowed for the SU-I scenario. Although there is a possibility of \( R_{F_H}^{\tau \mu} > R_{F_H}^{\tau \mu,\, \rm{SM}} \) for the S-V, S-IX, and S-XIII scenarios, the potential enhancement is only marginal. Therefore, it would be challenging to discern the nature of new physics through the measurement of \( R_{F_H}^{\tau \mu} \).

Thus, none of the basic observables in $B\to K \ell \ell$ decay can be utilized to construct a genuine LFUV observable in the $\tau -\mu$ sector.
However, using the definitions of $\Gamma (B \to K \ell^+ \ell^-)$ and $F^\ell_H $, one can construct the following observable
\begin{equation}
    \Gamma_\ell (1-F^\ell_H) = -\frac{4}{3}\int_{q^2_{\rm min}}^{{q^2_{max}}} dq^2 \,c_\ell(q^2)\,.
\end{equation}
It is obvious from the right-hand side of the above equation that $m_\ell$ doesn't appear in any term except in factor $\beta_\ell$, which appears as a common factor to all WCs. Therefore, the LFU ratio utilizing the observable  $  \Gamma_\ell (1-F^\ell_H)$, in principle,  can be termed as the genuine LFU ratio to test LFUV new physics in $\tau -\mu$ sector in $B\to K \ell \ell$ decays.

On the experimental front, the investigation of \( b \to s \tau^+ \tau^- \) transitions is currently hindered by the technical challenges involved in the reconstruction of tau leptons in decay products, resulting in only upper limits being available for these processes. These limits are significantly above the predictions made by the SM. For example, the measured upper bounds on the branching ratios for \( B \to K \tau^+ \tau^- \) and \( B \to K^* \tau^+ \tau^- \) are set at \(2.25 \times 10^{-3}\) \cite{BaBar:2016wgb} and \(2 \times 10^{-3}\) \cite{Belle:2021ndr}, respectively.

Addressing this challenge is crucial for advancing our understanding of potential new physics phenomena through these and similar decays, such as $b \to d \tau^+ \tau^-$ \cite{Belle-II:2018jsg,Alonso:2015sja,Capdevila:2017iqn,Ali:2023kvz,Li:2023mrj,Panda:2024ygr,Bordone:2023ybl}. Achieving significant improvements in tau-reconstruction technology is therefore imperative. Such advancements are expected to be realized at state-of-the-art experimental facilities, including the High-Luminosity LHC (HL-LHC) \cite{LHCb:2018roe}, Belle II \cite{Belle-II:2018jsg}, and the Future Circular Collider in electron-positron mode (FCC-ee) \cite{Bernardi:2022hny,Kamenik:2017ghi,Li:2020bvr}.
Current projections indicate that the HL-LHC and Belle II could potentially detect $B \to K \tau^+ \tau^-$ and  $B \to K^* \tau^+ \tau^-$ decays with improved sensitivities in the range of  $10^{-4}$ to $10^{-5}$, pushing the limits of detection closer to those expected by the SM. Furthermore, the FCC-ee, with its advanced vertex reconstruction capabilities, is poised to not only accurately measure the branching ratios at SM levels but also to provide a detailed analysis of the angular distributions of these decays.

\section{Conclusion}
\label{conc-gen}
In this study, we have extensively investigated the behavior of various LFU ratios in $b \to s \ell \ell$ decays to discern genuine signatures of LFUV within the $\tau-\mu$ sector. Our analysis builds upon previous findings that unlike the well-studied ratios $R_K^{\mu e} \equiv R_K \equiv \Gamma(B \to K \mu^+ \mu^-)/\Gamma(B \to K e^+ e^-)$ and $R_{K^*}^{\mu e} \equiv R_{K^*} \equiv \Gamma(B \to K^* \mu^+ \mu^-)/\Gamma(B \to K^* e^+ e^-)$,  the $R_K^{\tau \mu}$ and $R_{K^*}^{\tau \mu}$ ratios may exhibit deviations even under scenarios involving universal new physics couplings to leptons. This observation underscores the necessity to delineate and validate genuine LFUV observables within the $\tau-\mu$ sector. 

We focus on the analysis of the full angular distributions in the decays $B \to K \ell \ell$ and $B \to K^* \ell \ell$, aiming to identify robust LFUV observables. In the context of $B \to K^* \ell \ell$ decays, we find that:
\begin{itemize}
    \item  Analogous to $R_{K^*}^{\tau \mu}$, the ratios $R_{A_{FB}}^{\tau \mu}$ and $R_{f_L}^{\tau \mu}$ do not meet the criteria for genuine LFUV observables due to their sensitivity to mass effects.
    \item In contrast,  all optimized observables in the $B \to K^* \ell \ell$ decays within the $\tau-\mu$ sector are genuine LFUV observables, providing a reliable framework for probing LFUV new physics.
\end{itemize}

For $B \to K \ell \ell$ decays, 
\begin{itemize}
    \item the ratio $R_{F_H}$, similarly to $R_K^{\tau \mu}$, is affected by mass effects and thus fails to qualify as a genuine LFUV observable.
    \item we construct the ratio $\Gamma_\tau(1-F_{H}^{\tau})/\Gamma_\mu(1-F_{H}^{\mu})$ which can serves as the sole genuine LFUV observable for $B \to K \ell \ell$ decays involving tau and muon leptons.
\end{itemize}

 Moreover, leveraging new physics Lorentz structures that provide a better fit to  $b \to s \ell \ell$ data as compared to the SM, our study also elucidates how ratios that do not qualify as genuine LFUV observables—such as $R_{A_{FB}}^{\tau \mu}$ and $R_{f_L}^{\tau \mu}$—can nonetheless be instrumental in distinguishing between models featuring exclusively universal lepton couplings and those incorporating both universal and non-universal couplings. 

\section{Acknowledgement}
{\small We thank the anonymous referees for their valuable suggestions, which have enhanced the quality of this work. The work of JK is supported by SERB-India Grant EEQ/2023/000959. AKA would like to thank Shireen Gangal for useful suggestions and discussions.}

\end{document}